\documentclass[prd,twocolumn,floats,floatfix,nofootinbib]{revtex4}
\usepackage[dvips]{graphicx}
\usepackage{graphics}
\usepackage{dcolumn}
\usepackage{amssymb}
\usepackage{bm}
\usepackage{amsmath}
\usepackage{color}

\newcommand{\mS}{{\cal S}}

\begin{document}

\title{Bouncing Bianchi Type I Universes with Interacting Vacuum}
\author{Marco Bruni${{^{ab}}}$\footnote{marco.bruni@port.ac.uk}, Rodrigo Maier${{^c}}$\footnote{rodrigo.maier@uerj.br},
Thomas Waters${{^d}}$\footnote{thomas.waters@port.ac.uk}
\vspace{0.5cm}}

\affiliation{${{^a}}$Institute of Cosmology {\rm \&} Gravitation, University of Portsmouth, Dennis Sciama Building, Burnaby Road, Portsmouth, PO1 3FX, United Kingdom\\
\\
${{^b}}$INFN Sezione di Trieste, Via Valerio 2, 34127 Trieste, Italy\\
\\
${{^c}}$Departamento de F\'isica Te\'orica, Instituto de F\'isica, Universidade do Estado do Rio de Janeiro,\\
Rua S\~ao Francisco Xavier 524, Maracan\~a,\\
CEP20550-900, Rio de Janeiro, Brasil\\
\\
${{^d}}$School of Computing, Mathematics and Physics, University of Portsmouth, Lion Gate Building,\\
Lion Terrace, Portsmouth, Hampshire PO1 3HF, UK
}


\date{\today}

\begin{abstract}
We investigate the global dynamics and cosmological viability of non-singular 
solutions in an anisotropic Bianchi type I universe featuring an interacting vacuum 
energy component $V$ and a perfect fluid $\rho$, extending previous work based on a Robertson-Walker geometry. By modeling the energy-momentum 
transfer via a phenomenological four-vector $Q_\mu$, we reduce the field equations to 
a four-dimensional autonomous dynamical system for the expansion rate $\Theta$, total 
shear $\sigma$, energy density $\rho$, and vacuum component $V$. We examine two 
distinct forms for how the interaction depends on $\rho$ and $V$: a linear coupling $Q_1 \propto \Theta[\xi(V_\Lambda - V) 
+ \chi \rho]$ and a non-linear coupling $Q_2 \propto \Theta \rho (1 - V/V_\Lambda)$. 
Using a potential-energy analogy and phase-space analysis, we show that the linear 
model $Q_1$ naturally admits non-singular bouncing trajectories. In contrast, unlike the Robertson-Walker case,  for the 
non-linear interaction $Q_2$, invariant boundary surfaces in phase space strictly 
constrain the domain of expansion ruling out cosmological bounce scenarios for 
physical matter configurations. 
\end{abstract}
\maketitle

\section{Introduction}

The standard cosmological model, based on General Relativity and the FLRW metric, has 
proven remarkably successful in describing the large-scale evolution and 
observational features of our Universe \cite{Planck:2018vyg, 
SupernovaSearchTeam:1998fmf}. Despite its success, the paradigm suffers from deep 
conceptual challenges at both temporal extremes. At early times, standard FLRW 
cosmologies inevitably predict an initial cosmological singularity -- a state of 
infinite energy density, temperature, and spacetime curvature where the classical 
laws of physics break down \cite{Hawking:1970zqf,Hawking:1973uf,Wald:1984rg}. At late times, explaining the 
observed accelerated expansion requires a dark energy component, most simply modeled 
as a cosmological constant $\Lambda$, which suffers from extreme fine-tuning and 
coincidence problems \cite{Weinberg:1988cp, Peebles:2002gy}.

To overcome the initial singularity problem, non-singular bouncing cosmologies have 
emerged as a compelling complement, or even a complete alternative, to standard inflationary scenarios 
\cite{Novello:2008ra, Brandenberger:2016vhg, Cai:2014bea}. In a bouncing scenario, 
the Universe undergoes a contracting phase prior to reaching a minimal non-zero 
spatial volume, after which it smoothly transitions into the present expanding epoch. 
Generating such a bounce within standard GR generally requires violating the Null 
Energy Condition (NEC) -- namely, $\rho + p \ge 0$ -- which implies the presence of
exotic matter fields  near the bounce radius \cite{Cai:2008qb}. Generating such a bounce within standard GR generally requires violating the Strong Energy Conditions 
(SEC) -- namely, for a perfect fluid, $\rho + 3 p \ge 0$, or even the  Null 
Energy Condition (NEC) -- namely, for a perfect fluid in a spatially flat FLRW cosmology, $\rho + p \ge 0$ -- which implies the presence of
exotic matter fields  near the bounce, see e.g.\  \cite{Cai:2008qb}.

Simultaneously, dynamical models where the vacuum energy density $V(t)$ interacts non-trivially with a 
matter fluid $\rho$ offer a compelling framework to address late-time cosmological tensions, such as the $H_0$ 
discrepancy, while alleviating the dark energy coincidence problem 
\cite{Salvatelli:2014zta,Benetti:2021div,Wang:2015wga,Zhao:2017cud,Sola:2016sbt,DiValentino:2017iww,Martinelli:2019dau,Hogg:2020rdp,Hogg:2021yiz,Sebastianutti:2023dbt}. More 
generally, interacting vacuum scenarios preserving the vacuum equation of state $w = -1$ have been 
extensively investigated \cite{Bertolami:1986bg,Alcaniz:2012mh,Freese:1986dd,Carvalho:1991ut,Shapiro:2000dz,Wands:2012vg}, 
with significant theoretical motivation arising from quantum field theory in curved spacetime -- most 
notably through the Running Vacuum Model framework, where $V(t)$ evolves as a series of even powers of 
the Hubble parameter \cite{Chen:1990jw,Lima:2013dmf,Moreno-Pulido:2020anb}.

Coupling such dynamic vacuum states to matter allows for a continuous energy-momentum exchange mediated by a phenomenological interaction $4$-vector $Q_\mu$, a mechanism which has been shown to give rise to non-singular cosmological solutions \cite{Bruni:2021msx}. Crucially, this energy transfer modifies the effective equation of state of the composite fluid, enabling non-trivial early- and late-time dynamics. Beyond cosmology, such interactions have proven influential in gravitational 
collapse scenarios, where energy exchange between non-relativistic matter and a vacuum component can modify 
end-state geometries, such as driving collapsing dust or radiation toward regular or Reissner-Nordstr\"om-de 
Sitter configurations \cite{Maier:2020bgm}. This versatility suggests that interacting vacuum energy may 
naturally alter the overall energy-momentum tensor near high-density regimes, potentially triggering non-singular bounces. More generally, bouncing cosmologies have been shown to arise from models where the dark energy component, which may be interacting, has an equation of state other than the vacuum case, i.e. $w\not = -1$ \cite{Burkmar:2023tll,Burkmar:2024qpu,Burkmar:2025svw}. 

However, a realistic cosmological model must also account for spatial anisotropies,
which dominate the dynamics of the early Universe as spatial volume contracts 
\cite{Bozza:2009jx,Misner:1969hg}. In anisotropic models such as the Bianchi spacetimes, unless a negative anisotropic pressure takes over at high energies (see e.g.\ \cite{Ganguly:2019llh}), the 
shear energy density scales as $\sigma^2 \propto a^{-6}$ for a perfect fluid, growing rapidly during 
contraction and often causing an anisotropy that destabilizes the bounce 
or leads to a chaotic Belinski-Khalatnikov-Lifshitz (BKL) singularity 
\cite{Belinsky:1970ew,Wainwright:1997dsc}. Consequently, establishing whether an 
interacting dynamic vacuum can suppress or accommodate anisotropic shear while 
providing a stable, non-singular bounce remains a critical open question \cite{Bozza:2009jx}.

In this work, we conduct a global phase-space and dynamical systems analysis 
of an anisotropic Bianchi type I universe featuring an interacting vacuum 
component $V$ and a perfect fluid $\rho$, generalizing the isotropic case analyzed in \cite{Bruni:2021msx}. By modeling the interaction through
two distinct phenomenological couplings -- a linear coupling 
$Q_1 \propto \Theta[\xi(V_\Lambda - V) + \chi \rho]$ and a non-linear coupling
$Q_2 
\propto \Theta \rho (1 - V/V_\Lambda)$ -- we 
map out the complete cosmological dynamics. Using a potential-energy 
formulation and 
phase-space reduction, we demonstrate that while the linear interaction $Q_1$ 
naturally supports non-singular bouncing trajectories across an anisotropic 
background, the non-linear coupling $Q_2$ introduces invariant boundaries that 
strictly forbid cosmological bounces for physical matter configurations, unlike the isotropic case.

The paper is organized as follows: In Section II, we set up the Einstein field 
equations and conservation laws for an anisotropic Bianchi I spacetime with 
energy-momentum transfer, casting the equations into a closed four-dimensional
autonomous system. Section III focuses on the linear interaction model $Q_1$,
developing an 
effective potential analogy, analyzing phase-space orbits, and detailing the
NEC-violating conditions near the bounce point. Section IV presents the global
phase dynamics of the non-linear coupling $Q_2$ and proves the strict absence
of physical bouncing solutions. Finally, we summarize our findings and discuss
their cosmological implications in Section V.

\section{Theoretical Framework}

We start by considering the field equations
\begin{eqnarray}
\label{eq2}
G_{\mu\nu}= \kappa^2 (T_{\mu\nu}-V g_{\mu\nu})
\end{eqnarray}
where $G_{\mu\nu}$ is the Einstein tensor and $T_{\mu\nu}$ is the energy momentum tensor which comes from a perfect fluid 
\begin{eqnarray}
\label{eq3}
T^{\mu}_{~~\nu}&=&(\rho+p)u^\mu u_\nu+ p \delta^{\mu}_{~\nu}
\end{eqnarray}
which interacts with the vacuum component $V$ through a $4$-vector $Q_\nu$. Throughout this work, we adopt natural units such that $\kappa^2 \equiv 1$.

From Bianchi identities it follows that
\begin{eqnarray}
\label{eq4}
\nabla_\mu (T^\mu_{~\nu})=-Q_\nu,\\
\label{eq4p}
-\nabla_\nu V=Q_\nu.
\end{eqnarray}
In the following we are going to restrict ourselves to the case
\begin{eqnarray}
\label{eq10}
Q^\mu =Q u^\mu. 
\end{eqnarray}

Let us consider the Bianchi type I metric whose line element in the coordinates $(t, x, y, z)$ can be written as
\begin{eqnarray}
ds^2=-dt^2+a^2(t)dx^2+b^2(t)dy^2+c^2(t)dz^2.
\end{eqnarray}
Fixing $u^\mu=\delta^\mu_{~t}$, equations (\ref{eq4}) and (\ref{eq4p})
turn into
\begin{eqnarray}
\label{eq11p}
\dot{\rho}+\Big(\frac{\dot{a}}{a}+\frac{\dot{b}}{b}+\frac{\dot{c}}{c}\Big)(\rho+p)&=&-Q,\\
\label{eq111}
\dot{V}&=&Q.
\end{eqnarray}

From Einstein field equations (\ref{eq2}) on the other hand, it follows:
\begin{eqnarray}
\label{eq00}
&&\frac{\dot{a}\dot{b}}{ab}+\frac{\dot{b}\dot{c}}{bc}+\frac{\dot{a}\dot{c}}{ac}=(\rho+V),\\
\label{eq11}
&&\frac{\ddot{b}}{b}+\frac{\ddot{c}}{c}+\frac{\dot{b}\dot{c}}{bc}=-(p-V),\\
\label{eq22}
&&\frac{\ddot{a}}{a}+\frac{\ddot{c}}{c}+\frac{\dot{a}\dot{c}}{ac}=-(p-V),\\
\label{eq33}
&&\frac{\ddot{a}}{a}+\frac{\ddot{b}}{b}+\frac{\dot{a}\dot{b}}{ab}=-(p-V).
\end{eqnarray}

Defining  
\begin{eqnarray}
\label{eqex}
\Theta&=&\nabla_\alpha u^\alpha\equiv
\frac{\dot{a}}{a}+\frac{\dot{b}}{b}+\frac{\dot{c}}{c},\\
\sigma_{\alpha\beta}&=&\nabla_{(\alpha} u_{\beta)}-\frac{1}{3}\Theta h_{\alpha\beta}+\dot{u}_{(\alpha}u_{\beta)},
\end{eqnarray}
where $h_{\alpha\beta}=g_{\alpha\beta}+u_\alpha u_\beta$ is the projector on spatial surfaces orthogonal to $u^\alpha$,
it can be shown that
(\ref{eq00}) can be rewritten as 
\begin{equation}
\label{eq4nnn}
    \frac13 \Theta^2 - \sigma^2  - (\rho+V) = 0 \,,
\end{equation}
where, $\sigma^2\equiv  \frac{1}{2}\sigma_{ij}\sigma^{ij} $ is the shear, namely,
\begin{eqnarray}
\label{eqs2}
    \sigma^2  
    = \frac{1}{3} \Big(\frac{\dot{a}^2}{a^2}+\frac{\dot{b}^2}{b^2}+\frac{\dot{c}^2}{c^2}-\frac{\dot{a}\dot{b}}{ab}-\frac{\dot{b}\dot{c}}{bc}-\frac{\dot{a}\dot{c}}{ac}\Big). 
\end{eqnarray}
From (\ref{eq11})--(\ref{eq33}) on the other hand, we obtain
\begin{eqnarray}
\label{eq1350}
\dot{\Theta}=-\frac{1}{2}\Theta^2-\frac{3}{2}\sigma^2+\frac{3}{2}(V-w\rho),    
\end{eqnarray}
where we have used the equation of state $p=w\rho$.
From (\ref{eq4nnn}) and (\ref{eq1350}) one may obtain the 
Raychaudhuri equation in its standard from:
\begin{eqnarray}
\label{rac}
\dot{\Theta}=-\frac{1}{3}\Theta^2-2\sigma^2-\frac{1}{2}[\rho(1+3w)-2V].    
\end{eqnarray}

Defining
\begin{eqnarray}
\sigma_+&=&\frac{1}{2}(\sigma^{y}_{~y}+\sigma^{z}_{~z})=\frac{1}{6}\Big(\frac{\dot{b}}{b}+\frac{\dot{c}}{c}-\frac{2\dot{a}}{a}\Big),\\
\sigma_-&=&\frac{1}{2\sqrt{3}}(\sigma^{y}_{~y}-\sigma^{z}_{~z})=\frac{1}{2\sqrt{3}}\Big(\frac{\dot{b}}{b}-\frac{\dot{c}}{c}\Big),
\end{eqnarray}
we introduce two dynamical quantities which measure different types of spatial anisotropies. As defined above $\sigma_+$ measures the deformation of the $x$-axis relative to the $y$-$z$ plane while $\sigma_-$ measures the shape distortion within the $y$-$z$ plane itself. Straightforward calculations show that
\begin{eqnarray}
\dot{\sigma}_\pm=-\Theta \sigma_\pm,    
\end{eqnarray}
where we have used (\ref{eq11})--(\ref{eq33}). It is worth noting that
as $\sigma^2=3(\sigma_+^2+\sigma_-^2)$ one can obtain an analogous dynamical equation for the total shear, namely
\begin{eqnarray}
\label{eds}
\dot{\sigma}=-\Theta \sigma.     
\end{eqnarray}

To summarize, we have the following 1st order system in the four variables $\Theta,\sigma,\rho,V$:

\begin{eqnarray}
\dot{\Theta}&=&-\frac{1}{3}\Theta^2-2\sigma^2-\frac{1}{2}[\rho(1+3w)-2V],\label{4sys1} \\
    \dot{\sigma}&=&-\Theta \sigma,\label{4sys2} \\\dot{\rho}&=&-(1+w)\rho\Theta-Q, \label{4sys3} \\ 
\dot{V}&=&Q, \label{4sys4}
\end{eqnarray} along with the constraint \begin{eqnarray}
     \frac13 \Theta^2 - \sigma^2  - (\rho+V) = 0. \label{constorig}
\end{eqnarray}

In the following we shall examine two phenomenological covariant choices for the coupling $Q$, namely,
\begin{eqnarray}
\label{eq12}
Q_1&=&[\xi(V_\Lambda-V)+\chi\rho]\Theta,\\
Q_2&=&\zeta (1-V/V_\Lambda)\rho\Theta \label{quad}
\end{eqnarray} which we could describe as linear and quadratic (in $\rho$ and $V$) respectively.

\section{Nonsingular Solutions for $Q=Q_1$}

For the choice $Q=Q_1$, we can integrate
equations (\ref{4sys3}) and (\ref{4sys4}) by first defining a new independent variable $\ln(abc)$, whose $t$ derivative is $\Theta$, to give a pair of linear ODE's whose solutions are
\begin{eqnarray}
\label{eqa81n}
\rho &=& E_1 (abc)^{\alpha_1/3}+E_2 (abc)^{\alpha_2/3},\\
\label{eqa82n}
V&=&V_\Lambda+\lambda_1 (abc)^{\alpha_1/3}+\lambda_2 (abc)^{\alpha_2/3},
\end{eqnarray}
where
\begin{eqnarray}
\label{eqa91}
\alpha_1&=&-\frac{3}{2}(1+w+\xi + \chi+\Delta),\\
\label{eqa92}
\alpha_2&=&-\frac{3}{2}(1+w+\xi + \chi-\Delta),
\end{eqnarray}
with
\begin{eqnarray}
\label{delta}
\Delta=\sqrt{(1+w+\chi-\xi)^2+4\chi\xi}.
\end{eqnarray} We suppose $\alpha_1,\alpha_2$ to be real and distinct, and we note $\alpha_1<\alpha_2$. The constants of integration $\lambda_1$, $\lambda_2$, $E_1$ and $E_2$ are subject to the constraints 
\begin{eqnarray}
\label{eqa101}
 \lambda_1 \xi - E_1 \Big(1+w+\chi+\frac{\alpha_1}{3}\Big)=0,\\
\label{eqa102}
\lambda_2 \xi - E_2 \Big(1+w+\chi+\frac{\alpha_2}{3}\Big)=0.
\end{eqnarray}

It can be convenient to recast our equations in terms of dimensionless variables; if we assume that we have an energy scale in our system, $\rho_*$, we may define 
\begin{eqnarray}
\label{redef}
y=\frac{\Theta}{\sqrt{\rho_*}},~~z=\frac{\sigma}{\sqrt{\rho_*}},~~x=\frac{{\rho}}{\rho_*},~~v=\frac{V}{\rho_*}.
\end{eqnarray} and a time rescaling $\tau=\sqrt{\rho_*}t$. 
In this case the constraint (\ref{constorig}) can be written as
\begin{eqnarray}
\label{eq28}
y^2=3(x+v+z^2)    
\end{eqnarray}
and the system (\ref{4sys1})-(\ref{4sys4})
becomes (with $Q=Q_1$)
\begin{eqnarray}
\label{eqypn}
&&y^\prime=-\frac{1}{2}\Big[x(1+3w)+\frac{2y^2}{3}+4z^2-2v\Big]\\
\label{sheareq}
&&z^\prime=-yz\\
\label{dsb11}
&&x^\prime=-y[x(1+w+\chi)-\xi(v-v_\Lambda         )]\\
\label{dsb14}
&&v^\prime=y[\chi x-\xi(v - v_\Lambda)   ].
\label{vpnw}
\end{eqnarray}
where we have defined $v_\Lambda\equiv V_\Lambda/\rho_*$.

We will find that the average scalar factor 
\begin{eqnarray}
{\cal S}=(abc)^{\frac{1}{3}},    
\end{eqnarray}
can be a useful variable to understand the dynamical evolution of our system (note $\mS\geq 0$); for example we see that equation (\ref{sheareq}) can be easily integrated furnishing

\begin{eqnarray}
\label{solz}
z=\frac{z_0}{\sqrt{\rho_*}}(abc)^{-1}= \frac{z_0}{\sqrt{\rho_*}\mS^3}.   
\end{eqnarray}

\subsection{Potential energy analogy}

With the use of (\ref{eqa81n}), ({\ref{eqa82n}}), (\ref{solz}) and (\ref{redef}), equation (\ref{eq28}) can be rewritten as
\begin{eqnarray}
y^2+U({\cal S})=1, \label{energy}  
\end{eqnarray}
which takes the form `kinetic plus potential' since $y=3{\cal S}'/{\cal S}$, and we have fixed $\rho_*=3V_\Lambda$ (and hence $v_\Lambda=1/3$) so that the `potential energy' is
\begin{eqnarray}
\label{gueff}
U({\cal S})=-\frac{1}{V_\Lambda}\Big[(E_1+\lambda_1){\cal S}^{\alpha_1}+(E_2+\lambda_2){\cal S}^{\alpha_2}+\frac{z_0^2}{{\cal S}^6}\Big].\,    
\end{eqnarray}
As we are primarily interested in bouncing or non-singular cosmologies (i.e.\ potential barriers which stop ${\cal S}$ from reaching zero) we will first consider the case where the potential has an infinite barrier and examine the behaviour of $\rho$ and $V$ as ${\cal S}$ approaches the bounce.

Firstly we write (\ref{eqa81n}), ({\ref{eqa82n}}) as \begin{align} 
\label{linsol}
\rho(\mathcal{S})&=E_1\mathcal{S}^{\alpha_1}+E_2\mS^{\alpha_2},\\ V(\mS)&=V_{\Lambda}+\lambda_1\mS^{\alpha_1}+\lambda_2\mS^{\alpha_2}. \end{align} As $\mS\to+\infty$ we would expect $\rho\to 0$ so from now on we consider $\alpha_1,\alpha_2<0$, in which case \begin{align} \lim_{\mS\to+\infty}(\rho,V)=(0,V_{\Lambda})\label{asym} \end{align} where we suppose $V_{\Lambda}>0$, and since $\alpha_1\alpha_2=9\xi(1+w)$ and $w>0$ then $\xi>0$. We also note that \begin{align} \frac{\mS}{3}\frac{d}{d\mS}(\rho+V)=-(1+w)\rho.\label{rpv} \end{align}

\begin{figure}[tbp]
\includegraphics[width=8cm,height=5cm]{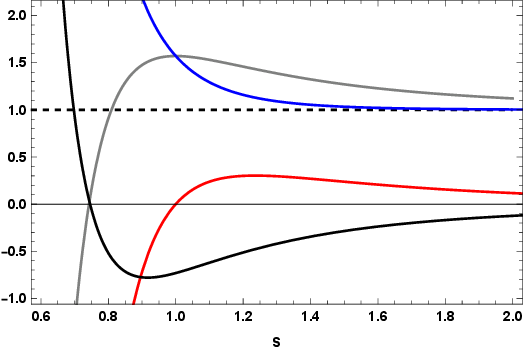}
\caption{The typical behaviour of the potential (black curve), energy density (red curve), vacuum component (blue curve) and combined density $\rho + V$ (gray) profiles as a function of the average scale factor $\mS$ in the case of an infinite potential barrier. In this numerical illustration we fixed $w=0$, $\xi=7/3$, $\chi=0$ so that $\alpha_1=-7$ and $\alpha_2=-3$.
Other parameters are fixed to the values $E_1=-1.0$, $E_2=1.0$, $z_0=0.4$ and $V_\Lambda=1.0$. The dashed black horizontal line indicates the total energy of the system according to (\ref{energy}). As described in the text, both the total energy density $\rho$ and the effective energy density $\rho + V$ become negative as the system approaches the bounce.
}
\label{fig1}
\end{figure}

The potential energy can be written as 
%
\begin{eqnarray}
U=\frac{1}{V_\Lambda}\Big[3(1+w)\left(\frac{E_1}{\alpha_1}S^{\alpha_1}+\frac{E_2}{\alpha_2}S^{\alpha_2}\right)-\frac{z_0^2}{S^6}  
\Big]
\end{eqnarray}
and for an infinite potential barrier to exist, i.e.\ $U\to+\infty$ as $S\to 0^+$, we require 
$\alpha_1<-6$ and $E_1<0$. From these conditions we can sketch the shape of $\rho, V$ and $\rho+V$, see Figure \ref{fig1} for a typical example; in particular since $\rho$ should not be always negative then we require 
$E_2>0$ which from \eqref{linsol} means there is a special value of $\mS$, which we label $\mS_0$, where $\rho$ is negative for $\mS<\mS_0$ and positive for $\mS>\mS_0$. 
Since 
%
%
\begin{eqnarray}    
\frac{d U}{d \mS}\Big|_{\mS_0}=6 \frac{z_0^2}{\mS_0^7}>0
\end{eqnarray}
this means that if we have a potential barrier followed by a minimum, then $\mS = \mS_0$ will always be to the right of that minimum and therefore solutions reaching the barrier will necessarily stray into the $\mS < \mS_0$ zone, implying that $\rho$ is negative in a finite neighbourhood of the bounce.

Regarding the behaviour of $\rho+V$, from $\eqref{rpv}$ we see $\rho+V$ is increasing for $\mS<\mS_0$ and decreasing for $\mS>\mS_0$; indeed since \[ \rho+V=V_{\Lambda}-\frac{3E_1(1+w)}{\alpha_1}\mS^{\alpha_1}-\frac{3E_2(1+w)}{\alpha_2}\mS^{\alpha_2} \] then $\rho+V\to-\infty$ as $\mS\to 0^+$. This implies there is a value of $\mS$, which we label $\mS_1$, where $\rho+V$ becomes negative, and we want to determine if this is necessarily before the bounce.

Suppose the bounce happens at some value of $\mS$, which we label $\mS_2$; we can see from \eqref{energy} that $U(\mS_2)=1$. However, we know from \eqref{gueff} that \[U(\mS_1)=1-\frac{z_0^2}{V_\Lambda\mS_1^6}<U(\mS_2).\] Since $U$ has a potential barrier and is decreasing at the bounce, this means $\mS_2<\mS_1$, i.e.\ solutions will enter the $\rho+V<0$ domain before the bounce; see Figure \ref{fig1}.

The emergence of a negative energy density ($\rho < 0$) in the vicinity of the bounce point is a well known, albeit non-trivial, feature of many non-singular cosmological frameworks. This behavior is fundamentally rooted in the necessity of violating the Null Energy Condition (NEC), a requirement imposed by the Hawking-Penrose singularity theorems for any model seeking to avoid an initial singularity within the context of classical General Relativity\cite{Hawking:1970zqf}. In many bouncing scenarios, particularly those involving exotic matter or specific scalar field configurations like ghost condensates or quantum fields, the transition from contraction to expansion ($\dot{a} = 0, \ddot{a} > 0$) demands a stress-energy tensor that temporarily circumvents the Weak Energy Condition (WEC) to provide the necessary repulsive gravitational effect\cite{Cai:2008qb,Novello:2008ra}. While such violations raise questions regarding vacuum stability and the presence of ghost instabilities, they are often treated as effective descriptions of high-energy quantum corrections or modifications to the gravitational action that dominate at the Planck scale\cite{Brandenberger:2016vhg}.

\begin{center}
    *
\end{center}

More detail can be given in special cases; for example when $\chi=0$ we obtain
\begin{eqnarray}
\alpha_1=-\frac{3}{2}(1+w+\xi+|1+w-\xi|),\\
\alpha_2=-\frac{3}{2}(1+w+\xi-|1+w-\xi|).
\end{eqnarray}
For $1+w>\xi$, it is easy to show that
\begin{eqnarray}
\rho&=&\frac{E_1}{(abc)^{(1+w)}}+\frac{E_2}{(abc)^\xi},\\
V&=&V_\Lambda+\frac{E_2(1+w-\xi)}{\xi(abc)^\xi}.
\end{eqnarray}
so that
\begin{eqnarray}
\label{p1}
U({\cal S})=-\frac{1}{V_\Lambda}\Big[\frac{E_1}{{\cal S}^{3(1+w)}}+\frac{E_2}{\xi}\frac{(1+w)}{{\cal S}^{3\xi}}+\frac{z_0^2}{{\cal S}^6}\Big].  
\end{eqnarray}
Conversely, for $1+w<\xi$ we obtain
\begin{eqnarray}
\rho&=&\frac{E_1}{(abc)^{\xi}}+\frac{E_2}{(abc)^{(1+w)}},\\
V&=&V_\Lambda+\frac{E_1(1+w-\xi)}{\xi(abc)^\xi}.
\end{eqnarray}
so that
\begin{eqnarray}
\label{p2}
U({\cal S})=-\frac{1}{V_\Lambda}\Big[\frac{E_1}{\xi}\frac{(1+w)}{{\cal S}^{3\xi}}+\frac{E_2}{{\cal S}^{3(1+w)}}+\frac{z_0^2}{{\cal S}^6}\Big].  
\end{eqnarray}
As $0\leq w \leq 1$, one might see from the above that an infinite potential barrier 
-- avoiding the classical singularity -- 
may be obtained for the case in which
\begin{eqnarray}
\label{bd}
\xi> 2~~{\rm and}~~E_1<0.
\end{eqnarray}
In the absence of a stiff fluid, that is for $w<1$, it can be easily shown that nonsingular solutions may be obtained for $\xi=2$ as long as
\begin{eqnarray}
\label{bdn}
E_1<-\frac{2z_0^2}{(1+w)}.
\end{eqnarray}

\begin{figure}[tbp]
\includegraphics[width=8cm,height=5cm]{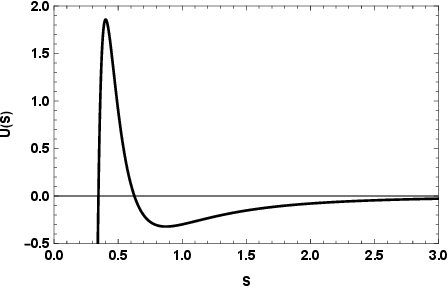}
\caption{The potential $U(\mS)$ for the parameters $w=0$, $E_1=-1.00$, $E_2=1.00$, $z_0=0.22$, $\xi=4/3$ and $V_\Lambda=1.00$. Here we see that nonsingular configurations can be obtained once the potential $U(\mS)$ displays a positive global maximum at $\mS_m$, such that $U(\mS_m) > 1$. After a contracting phase trajectories reach a potential barrier before expanding toward the de Sitter attractor.}\label{finitebarrier}
\end{figure}

It is worth emphasizing that nonsingular configurations can be achieved for alternative domains differing from (\ref{bd}) or (\ref{bdn}). Indeed, given $E_1 < 0$ and $\xi < 2$, one can demonstrate that the potential $U(\mS)$ admits a positive global maximum at $\mS_m$, such that $U(\mS_m) \geq 1$. Under these conditions, after a contracting phase trajectories encounter a potential barrier before evolving toward the de Sitter attractor. In Fig. \ref{finitebarrier} we show an example for a dust component.

For $\xi=0$ on the other hand,
\begin{eqnarray}
\label{potchi0}
U({\cal S})=-\frac{1}{V_\Lambda}\Big\{\Big(\frac{E_1}{1+w+\chi}\Big)\Big[\frac{1+w}{{\cal S}^{3(1+w+\chi)}}\Big]+\frac{z_0^2}{{\cal S}^{6}}\Big\}.    
\end{eqnarray}
In this case one may notice that nonsingular models might be obtained
as long as
\begin{eqnarray}
\frac{E_1}{1+w+\chi}<0~~{\rm and}~~3(1+w+\chi)>6.    
\end{eqnarray}
It is worth noting however that 
\begin{eqnarray}
\rho=\frac{E_1}{{\cal S}^{3(1+w+\chi)}}    
\end{eqnarray}
so that in the case of a FLRW model with a dust component
the usual matter dominated era  
can not be achieved for $\chi \neq 0$\cite{Bruni:2021msx}.
%

\subsection{The Phase Space}

A limitation of the previous analysis is that we have lost track of the time dependence of the variables, and one way to recover that is to consider the phase space picture; we will consider two such flows. Firstly in order to illustrate the behaviour of the $\mS$ variable we consider the dynamical system
\begin{eqnarray}
\label{ds1}
y^\prime&=&-\frac{1}{2}\Big[x(1+3w)+\frac{2y^2}{3}+4z^2-2v\Big],\\
\label{ds2}
\mS^\prime&=&\frac{y\mS}{3},
\end{eqnarray}
where the energy scale $\rho_*$ shall be restored so that
\begin{eqnarray}
x&=&\frac{1}{\rho_*}(E_1 \mS^{\alpha_1}+E_2 \mS^{\alpha_2}),\\
\label{eq60}
z&=&\frac{z_0}{\sqrt{\rho_*}\mS^3},\\
v&=&\frac{1}{\rho_*}({V_\Lambda}+\lambda_1 \mS^{\alpha_1}+\lambda_2 \mS^{\alpha_2}).
\end{eqnarray}

From (\ref{ds1})--(\ref{ds2}) we notice that fixed points are given by $y=0$ together with the condition
\begin{eqnarray}
\label{fp}
x(1+3w)+4z^2-2v=0.    
\end{eqnarray}
Since $x$, $z$, and $v$ depend explicitly on $\mS$, the aforementioned constraint determines the location of the fixed points, $\mS_c$, in the phase space. In order to simplify our analysis, in the following we shall restrict ourselves to the case $\chi=w=0$ and $\xi=2$.
In this case, (\ref{fp}) reduces to
\begin{eqnarray}
P(s)={-2V_\Lambda s^2+2E_1+ E_2 s+4z_0^2}=0,    
\end{eqnarray}
where $s\equiv \mS^3$. From the above one may notice that the phase space allows for at most two fixed points as long as
the condition
\begin{eqnarray}
\label{c2}
2E_1+\frac{E_2^2}{8V_\Lambda}+4z_0^2>0
\end{eqnarray}
holds. 
\begin{figure}[tbp]
\includegraphics[width=8cm,height=5cm]{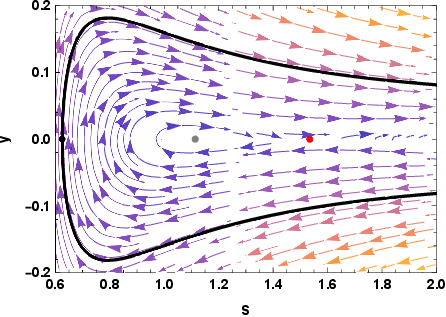}
\caption{Phase-space portrait of the dynamical system~(\ref{ds1})--(\ref{ds2})
for $\rho_*=100$, with $E_1=-1$, $E_2=1$, $z_0=0.5$,
$V_\Lambda=0.1$, and $\chi=w=0$, $\xi=2$.
The colored arrows represent the vector field of the unconstrained
two-dimensional dynamical system (\ref{ds1})--(\ref{ds2}) for selected initial conditions. The gray dot denotes the center at
$(\mS_1,0)$, with $\mS_1\simeq1.114$, while the red dot denotes the
saddle at $(\mS_2,0)$, with $\mS_2\simeq1.535$.
The black solid curve represent the physical cosmological trajectories
satisfying the Friedmann constraint (\ref{eq28}).
The bounce occurs at $\mS_b\simeq0.625$, where $y=0$ but
$y'\neq0$. The fixed points of the unconstrained dynamical system
do not lie on the Friedmann constraint and therefore do not represent
equilibrium cosmological solutions.
}
\label{fig3}
\end{figure}

To simultaneously satisfy \eqref{bdn} and \eqref{c2}, 
let us consider a numerical example with the 
parameters $E_1=-1$, $E_2=1$, $z_0=0.5$, 
$V_\Lambda=0.1$. Under these choices, the system 
yields two fixed points connected to
$S_1\simeq1.114$ and $S_2\simeq1.535$.
By defining the state vector
\begin{equation}
\Psi = \begin{bmatrix} y \\ \mS \end{bmatrix},
\end{equation}
we can linearize the system \eqref{ds1}--\eqref{ds2} to first order in the neighborhood of a fixed point $\Psi_c$ as
\begin{equation}
\Psi^\prime = L\big|_{\Psi_c}(\Psi - \Psi{c}),
\end{equation}
where the Jacobian matrix $L\big|_{\Psi_c}$ and the fixed point $\Psi_c$ are given by
\begin{equation}
L\big|_{\Psi_c} = \begin{bmatrix}
\frac{\partial y^\prime}{\partial y} & \frac{\partial y^\prime}{\partial \mS} \\
\frac{\partial \mS^\prime}{\partial y} & \frac{\partial \mS^\prime}{\partial \mS}
\end{bmatrix}\Big|_{\Psi_c}\quad 
\text{and} 
\quad\Psi_{c} = \begin{bmatrix} 0 \\ \mS_c \end{bmatrix}.
\end{equation}
Using the fixed-point polynomial
$P(s)$, one obtains
\begin{equation}
 \lambda^2=-\frac{P'(s_c)}{2\rho_*s_c},
 \qquad
 P'(s)=E_2-4V_\Lambda s.
\end{equation}
Hence the smaller root has $\lambda^2<0$ and is a center in the
linearized system, whereas the larger root has $\lambda^2>0$ and is a
saddle.

It is important, however, to distinguish the fixed points of this
two-dimensional extension from the physical cosmological trajectories.
The latter must also satisfy the Friedmann constraint
\begin{equation}
 y^2=3(x+v+z^2).
\end{equation}
For the present parameter choice this constraint can be written as
\begin{equation}
 y^2=\frac{3}{\rho_*}
 \left[
 V_\Lambda+\frac{E_2}{\mS^3}
 +\frac{E_1/2+z_0^2}{\mS^6}
 \right].
\end{equation}
The two fixed points above do not lie on this physical constraint
surface. Consequently, the center and saddle describe the phase
structure of the unconstrained $(\mS,y)$ system, but they are not
equilibrium cosmological solutions of the constrained Einstein system.

The physical turning point is instead obtained by imposing $y=0$ on
the Friedmann constraint:
\begin{equation}
 V_\Lambda \mS^6+E_2\mS^3+\frac{E_1}{2}+z_0^2=0.
\end{equation}
For the numerical example this gives $ \mS_b\simeq0.625$.
At this point $\mS'=0$ while $y'\neq0$, so it represents a bounce rather
than a fixed point.

Accordingly, the phase portrait should display both structures
separately: the full vector field of the two-dimensional system, which
contains the linear center and saddle, and the Friedmann constraint
curve, on which the physical bouncing solutions evolve. This interpretation is illustrated in Fig. 3. 
The arrows represent integral curves of the unconstrained dynamical system (\ref{ds1})--(\ref{ds2}) for fixed $(\rho_*=100)$, obtained from different initial conditions in the $(\mS,y)$ plane. The black curve identifies the physical cosmological trajectory selected by the Friedmann constraint (\ref{eq28}).
In particular, the physical
trajectories do not form closed cycles around the center because the
Friedmann constraint does not pass through either fixed point.
Qualitatively, rather than producing cyclic universes in the neighborhood of the center, the orbits remain open. Specifically, they emerge from a de Sitter repeller at infinity during the contraction phase, bounce within a finite neighborhood of the center, and subsequently begin their expansion toward a de Sitter attractor. 

Secondly, to illustrate the behaviour of the other variables we consider the dynamical system \eqref{eqypn}, \eqref{dsb11} and \eqref{dsb14}, removing $z^2$ with \eqref{eq28}, given by \begin{align*}
    x'&=-y[(1+w+\chi)x-\xi\tilde{v}], \\ \tilde{v}'&=y[\chi x-\xi \tilde{v}], \\ y'&=-\tfrac{1}{2}[-3(1-w)x-6\tilde{v}+2y^2-6v_\Lambda],
\end{align*} where $\tilde{v}=v-v_\Lambda$. There is much structure in this system: there is a symmetry $(\tau,y)\to(-\tau,-y)$ (such systems are known as reversible or $t$-invariant), which means every orbit in the upper half-space $y>0$ has a mirror image with the time arrow reversed in the lower half-space (in particular every orbit that intersects the $y=0$ plane does so perpendicularly); there are three invariant planes and a line of equilibria; but from the point of view of the current discussion the most important feature is that there are two equilibria at $(0,0,\pm\sqrt{3v_\Lambda})=(0,0,\pm 1)$ with eigenvalues $\pm(-2,\alpha_1,\alpha_2)$ which are therefore a repellor and an attractor (since $\alpha_1,\alpha_2<0$). We can see how the bouncing solution from the energy diagram in Figure \ref{fig1} fits into this 3D phase space in the following way:

\begin{figure}[h]
\includegraphics[width=0.5\textwidth]{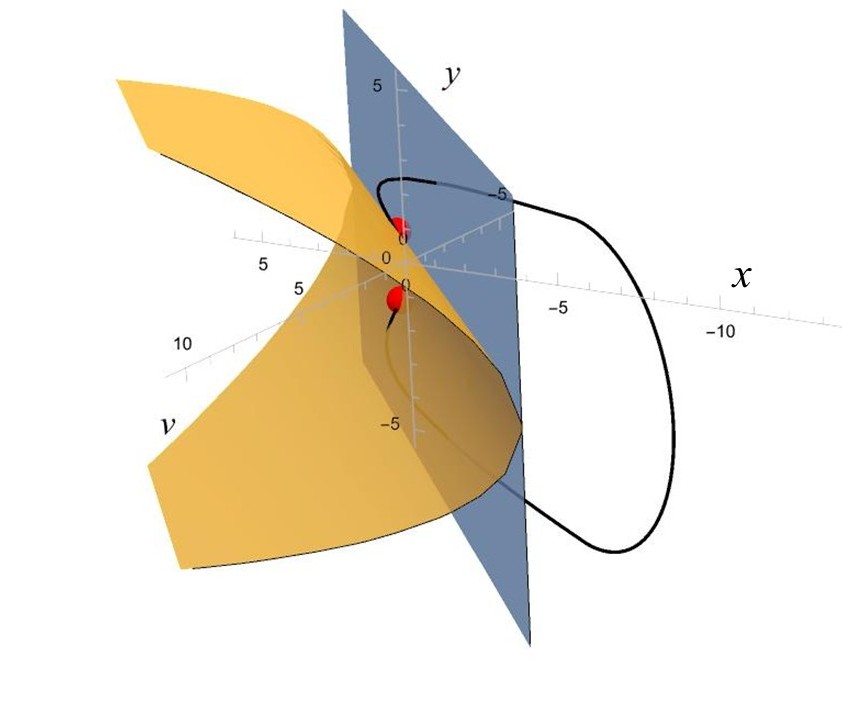}\caption{An orbit in the 3D $(x,v,y)$ space corresponding to a bouncing cosmology; the two red spheres are the de Sitter equilibrium points mentioned in the text, the trajectory in black asymptotes to these spheres in the past and future and crosses the $y=0$ plane perpendicularly.}\label{3Dphase}
\end{figure}




In the energy diagram suppose the $\mS$ value starts away out at $+\infty$ and starts decreasing; from \eqref{energy} that means $y=-1$ (since $\lim_{\mS\to+\infty}U=0$) and $\rho=0,V=V_{\Lambda}$ from \eqref{asym}; this is the $(x,\tilde{v},y)=(0,0,-1)$ repellor equilibrium point. $\mS$ decreases until the bounce ($\mS'=y=0$) and starts to increase again, now asymptoting to $(\rho,V,y)=(0,V_{\Lambda},+1)$. In the 3D phase space, this corresponds to an orbit which is heteroclinic (or doubly-asymptotic) to the two equilibria previously mentioned, the repellor in the past and the attractor in the future, and is symmetric about the $y=0$ plane. An example is shown in Figure \ref{3Dphase}, along with an invariant surface in orange (this is the $z^2=0$ surface so orbits cannot go into the convex part) and the $\rho+V=0$ plane in blue. Since those two surfaces touch one another along a line, the only way an orbit can go from the equilibrium point in the lower half-space to the equilibrium point in the upper half-space is by crossing the blue plane, and therefore passing into the $\rho+V<0$ region.

\section{The Nonlinear Coupling $Q=Q_2$}

We now turn our attention to the case of nonlinear coupling, with $Q=Q_2$ as in \eqref{quad}. In this case we obtain the following system of coupled equations
\begin{eqnarray}
\label{eqypnn}
&&y^\prime=-\frac{1}{2}\Big[x(1+3w)+\frac{2y^2}{3}+4z^2-2v\Big]\\
\label{sheareqn}
&&z^\prime=-yz\\
&&x^\prime=-[1+w+\zeta(1-v/v_\Lambda)]xy, \label{xvquadx}\\
\label{eq70}
&&v^\prime=\zeta(1-v/v_\Lambda)xy,\label{xvquadv}
\end{eqnarray}
subjected to the constraint
\begin{eqnarray}
\label{eq28nn}
y^2=3(x+v+z^2).    
\end{eqnarray}

Before a detailed analysis of the nature of the solutions we note that we are able to make some observations relating to bouncing cosmologies. According to equations (\ref{xvquadx}) and (\ref{eq70}) we see that $x=0$ and $v=v_\Lambda$ are invariant; this means that solutions that start in the region $x>0,v>v_\Lambda>0$ cannot leave it, and from the constraint equation \eqref{eq28nn} this implies that $y$ cannot be zero due to the presence of the quadratic shear term. As such there can be no bouncing cosmologies if at some time $x>0$ and $v>v_\Lambda$.

To thoroughly explore the dynamical landscape and fully understand the physical features of the solutions in this regime, we take a closer look of the system in the full three-dimensional phase space $(x, y, v)$ over the physical region $x \ge 0$, $v \ge 0$, and $0 \le w \le 1$. Utilizing the constraint \eqref{eq28nn} to eliminate $z$, the governing autonomous system reduces to:
\begin{align}
y^\prime &= -y^2+3v+\frac{3}{2}(1-w)x, \label{sys3d_y} \\
x^\prime &= - \left[ 1 + w + \zeta\left(1 - \frac{v}{v_\Lambda}\right) \right] x y, \label{sys3d_x} \\
v^\prime &= \zeta\left(1 - \frac{v}{v_\Lambda}\right) x y. \label{sys3d_v}
\end{align}
In the following we shall locate the equilibrium points $(y_c, x_c, v_c)$.
From the above we notice that fixing $y_c=0$ one obtains from (\ref{sys3d_y}) that:
\begin{eqnarray}
x_c=-\frac{2v_c}{1-w}    
\end{eqnarray}
which lies in the nonphysical domain for $v \ge 0$, and $0 \leq w \leq 1$. 

For $y_c\neq 0$ equation (\ref{sys3d_v}) admits two possible solutions, namely, $v_c=v_\Lambda>0$ or $x_c=0$. If $v_c = v_\Lambda$ the non-linear interaction term
decouples completely, while from (\ref{eq28nn}), (\ref{sys3d_y}) and (\ref{sys3d_x}) one obtains the algebraic relations 
\begin{eqnarray}
z_c=0,~~y_c^2 = 3v_\Lambda,~~{\rm and}~~x_c=0,    
\end{eqnarray}
respectively. Therefore, on the invariant boundary hypersurface $v = v_\Lambda$ we obtain a hyperbolic pair
of de Sitter equilibrium points: an expanding attractor $P_{\text{dS}}^+ = (0, +\sqrt{3v_\Lambda}, 
v_\Lambda)$ for $y > 0$, and a contracting repeller $P_{\text{dS}}^- = (0, -\sqrt{3v_\Lambda}, v_\Lambda)$ 
for $y < 0$. This very same result may be reproduced once one fixes $y_c\neq 0$ and $x_c=0$. 

From the above one may notice that our previous analysis covers the case $v\ge 0$ completely. Therefore, for 
$x>0$ there are no physical fixed points in the phase space. In fact if $x_c > 0$ and $y_c \neq 0$, one may 
note that enforcing $v^\prime = 0$ requires $v_c = v_\Lambda$, which in turn reduces \eqref{sys3d_x} to 
$x^\prime = -(1+w)xy$. For $0 \le w \le 1$, $x^\prime = 0$ can only be satisfied if $x_c = 0$, confirming the 
absence of fixed points with $x_c>0$.

The phase dynamics further clarify why cosmological bouncing solutions are forbidden in this domain. A 
cosmological bounce requires $y(t_0) = 0$ (a point of maximum contraction or minimum expansion). 
For any trajectory starting in the domain $x > 0$ and $v \ge 0$, the requirement $z^2 = \frac{1}{3}y^2 - x - 
v \ge 0$ enforces $y^2 \ge 3(x + v) > 0$. Thus, $y$ is strictly bounded away from zero by a finite positive 
gap $\sqrt{3(x+v)}$, physically preventing $y(t)$ from ever reaching $y = 0$
ruling out bouncing scenarios entirely.

\begin{figure}
\includegraphics[width=0.4\textwidth]{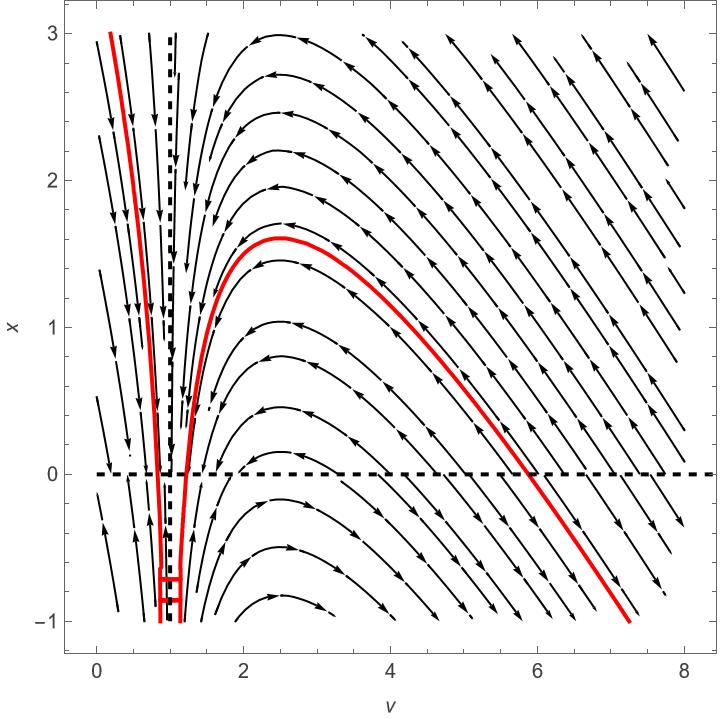}\caption{The $v$-$x$ phase plane according to \eqref{sys3d_x},\eqref{sys3d_v}, with $w=0.5,v_\Lambda=\zeta=1$; the red curve is $K_\Sigma=3.5$. Note the arrows are in the direction of increasing $\mS$.}\label{2Dphasevx}
\end{figure}

To provide a numerical visualization of the phase space, we eliminate $y$ from the coupled system (\ref{sys3d_x})--(\ref{sys3d_v}). Assuming $y \neq 0$, the reduced differential constraint reads
\begin{equation}
\label{edp}
x' + \left[ 1 + \frac{1 + w}{\zeta(1 - v/v_\Lambda)} \right] v' = 0.
\end{equation}
Integrating Eq.~(\ref{edp}) directly yields the conserved quantity
\begin{equation}
K_\Sigma = x + v - \frac{(1 + w)v_\Lambda}{\zeta} \ln \left| 1 - \frac{v}{v_\Lambda} \right|, \label{kseqn}
\end{equation}
where $K_\Sigma$ parameterizes a family of invariant surfaces $\Sigma$. Each surface consists of a union of phase curves, allowing us to map the system's trajectories directly onto the $(x, v)$ phase plane; an example is in Figure \ref{2Dphasevx}. These integral curves however do not give a sense of time dependence, so in equations (\ref{sys3d_x})--(\ref{sys3d_v}) we absorb $y$ as the independent variable as described before; now we have a two dimensional $x$-$v$ phase plane as in Figure \ref{2Dphasevx}, where the arrows are in the direction of increasing $\mathcal{S}$. Notice the horizontal dashed line $x=0$; the arrows point away from it on either side and so trajectories cannot cross this line. Similarly obits cannot cross the $v=v_\Lambda$ line, and so the region $x>0,v>v_\Lambda$ is invariant. In this region $y$ does not change sign (as explained above), and so the arrows are also in the direction of increasing time (when $y>0$).

\begin{figure}
\includegraphics[width=0.4\textwidth]{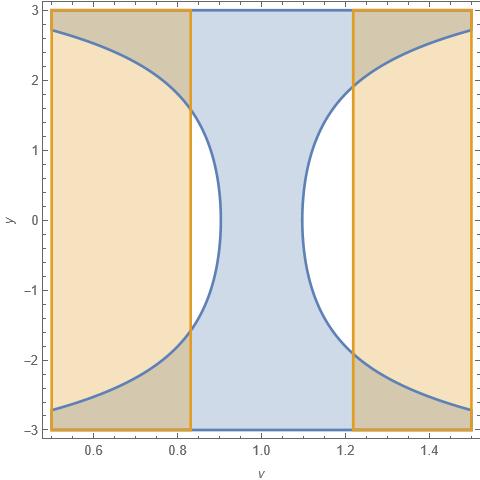}\caption{The $v$-$y$ phase plane with the same parameter values as Figure \ref{2Dphasevx}, with the regions $y^2/3-x(v)-v>0$ (blue) and $x(v)>0$ (orange) shaded. Note the overlap does not contain $y=0$, i.e.\ no bounce is possible in this model.}\label{2Dphasevy}
\end{figure}

Finally, we can use the explicit solution for $x=x(v)$ from \eqref{kseqn} to consider the two-dimensional $y$-$v$ phase space given by \eqref{sys3d_y},\eqref{sys3d_v}. Now the most relevant thing to observe is that the region $y^2/3-x(v)-v>0$ (required so $z^2>0$ by  \eqref{eq28nn}) and the region $x(v)>0$ (which is invariant) only overlap in disconnected components which do not contain the $y=0$ axis (see Figure \ref{2Dphasevy}); thus again we observe that trajectories that start with $x>0$ cannot lead to a bounce.

\section{Conclusions}

In this paper, we have presented a phase-space and dynamical systems analysis
of an anisotropic Bianchi type I universe containing a perfect fluid $\rho$ 
interacting with a dynamic vacuum component $V$. By modeling the energy-
momentum exchange through a phenomenological interaction $4$-vector $Q_\mu = Q 
u_\mu$, we reduced the field equations to a closed four-dimensional autonomous 
system in terms of the expansion scalar $\Theta$, total shear $\sigma$, fluid 
density $\rho$, and vacuum energy $V$. We investigated two distinct couplings: a linear interaction $Q_1 \propto \Theta[\xi(V_\Lambda - V) + \chi 
\rho]$ and a non-linear coupling $Q_2 \propto \Theta \rho (1 - V/V_\Lambda)$.

For the linear coupling $Q_1$, we established through a potential-energy 
analogy and phase-space reduction that the model naturally yields physically 
viable, non-singular bouncing cosmologies. Trajectories originating from a 
contracting de Sitter repeller at past infinity undergo a smooth bounce at a 
minimum average scale factor before expanding toward a late-time de Sitter
attractor. Near the bounce point, the system temporarily enters a region where 
both the matter density $\rho$ and the effective total energy density $\rho + 
V$ become negative, reflecting the localized violation of the Null Energy 
Condition necessary to bypass the classical singularity theorems
\cite{Hawking:1970zqf,Hawking:1973uf,Wald:1984rg}. 

In stark contrast, our global qualitative analysis of
the non-linear coupling $Q_2$ revealed that physical bouncing cosmologies are 
strictly forbidden. Invariant boundary hypersurfaces in the phase space -- 
specifically $x = 0$ and $v = v_\Lambda$ -- confine physical matter
trajectories such that the expansion rate scalar $y^2$ is bounded away from 
zero by a finite positive gap $3(x+v)$. 

The results obtained in this Bianchi type I framework open several compelling
avenues for future research. A natural next step is to extend this interacting
vacuum paradigm to more general anisotropic homogeneous spacetimes, most 
notably Bianchi type V  and Bianchi type IX cosmologies, as well as large-scale inhomogeneities around the possible bounce \cite{Ananda:2006gf,Ganguly:2019llh}.

In standard General Relativity, Bianchi V models represent open anisotropic 
universes, where spatial curvature interacts non-trivially with the expansion 
dynamics \cite{Ellis:1968vb}. Examining how linear ($Q_1$) and non-linear 
($Q_2$) interactions modify the shear dissipation rate and isotropization 
mechanism in Bianchi V spacetimes will reveal whether energy exchange can 
drive an anisotropic open universe toward spatial isotropy without fine-tuning 
\cite{Coley:2003mj}.

Furthermore, Bianchi type IX geometries provide the ultimate testing ground 
for early-universe dynamics due to their rich spatial curvature and chaotic 
behavior near the singularity. Contracting Bianchi IX models are notoriously dominated by chaotic, 
oscillatory Belinski-Khalatnikov-Lifshitz (BKL) bounces \cite{Maier:2015kma,Maier:2017dtb,Ganguly:2019llh}, where shear and 
spatial curvature terms repeatedly cause chaotic transitions between Kasner 
states \cite{Belinsky:1970ew,Wainwright:1997dsc}. Investigating whether $Q_1$-type 
linear energy-momentum transfer can suppress BKL oscillations and smooth out 
spatial curvature singularities -- effectively acting as an anisotropic dampening 
mechanism -- will provide key insights into whether dynamic vacuum interactions 
can offer a classical resolution to cosmic singularities across 
Bianchi class types.

\section{Acknowledgments}

RM acknowledges financial support from
FAPERJ Grant No. E-$26/010.002481/2019$.
For the purpose of open access, we have applied a Creative Commons Attribution (CC
BY) licence to any Author Accepted Manuscript version arising. Supporting research data
are available on reasonable request from the authors.

\end{document}